\documentclass[preprint,pre,showpacs,showkeys]{revtex4}%
\usepackage{amsfonts}
\usepackage{amsmath}
\usepackage{amssymb}
\usepackage{graphicx}%
\begin{document}
\preprint{Preprint }
\title[The anisotropic XY model on the inhomogeneous periodic chain]{Quantum transitions of the isotropic XY model with long-range interactions on
the inhomogeneous periodic chain }
\author{J. P. de Lima}
\affiliation{Departamento de F\'{\i}sica,Universidade Federal do Piau\'{\i}, Campus
Ministro Petr\^{o}nio Portela,64049-550, Teresina, Piau\'{\i}, Brazil.}
\author{L. L. Gon\c{c}alves}
\affiliation{Departamento de Engenharia Metal\'{u}rgica e de Materiais, Universidade
Federal do Cear\'{a}, Campus do Pici, Bloco 714, 60455-760, Fortaleza,
Cear\'{a}, Brazil}
\email{lindberg@fisica.ufc.br}
\author{}

\begin{abstract}
The isotropic XY model $(s=1/2)$\ in a transverse field, with uniform
long-range interactions among the transverse components of the spins, on the
inhomogeneous periodic chain, is studied. The model, composed of $N$\ segments
with $n$\ different exchange interactions and magnetic moments, is exactly
solved by introducing the integral gaussian transformation and the generalized
Jordan-Wigner transformation, which reduce the problem to the diagonalization
of a finite matrix of $n$th order. The quantum transitions induced by the
transverse field are determined by analyzing the induced magnetization of the
cell and the equation of state. The phase diagrams for the quantum
transitions, in the space generated by the transverse field and the
interaction parameters, are presented. As expected, the model presents
multiple, first- and second-order quantum transitions induced by the
transverse field, and it corresponds to an extension of the models recently
considered by the authors. Detailed results are also presented, at $T=0$, for
the induced magnetization and isothermal susceptibility $\chi_{T}^{zz}$ as
function of the transverse field.

\end{abstract}
\date{02/05/2008}

\pacs{05.70.Fh%
$\backslash$%
sep 05.70.Jk%
$\backslash$%
sep 75.10.Jm%
$\backslash$%
sep 75.10.Pq}
\keywords{Isotropic XY-model, long-range interaction, quantum transition,
one-dimensional model, inhomogeneous chain}\eid{identifier}
\maketitle

\section{Introduction}

The study of the critical quantum behavior of systems\cite{sachdev:2000},
which is induced by quantum fluctuations, has been object of \ great interest
in recent years. This critical behavior, which controls the properties of the
systems at very low temperature, is present in different systems. In
particular, for magnetic systems, they have been responsible for unusual
properties observed in low dimensional magnetic materials
\cite{gambardella:2002,mukherjeea:2004}. Therefore, the study of the critical
behavior of spin systems\ in low dimension, particularly the exactly soluble,
is of great importance for understanding the properties of these materials.

Among these models, the one-dimensional XY model introduced by Lieb, Schultz
and Mattis \cite{lieb:1961}, despite being almost fifty year old, is still the
best one to describe exactly magnetic quantum transitions. The rather rich
quantum critical behavior presented by the model can be seen in a recent work
by the authors \cite{delima:2007} and in the references therein, where they
study the anisotropic model on the inhomogeneous periodic lattice. The study
of the model on the inhomogeneous open lattice has also been recently
addressed by Feldman \cite{feldman:2006}.

The isotropic model on the inhomogeneous lattice has also been studied by the
authors \cite{delima:2006}, where a detailed study of the static and dynamic
critical properties is presented. Although the model has been applied mostly
in the study of the quantum\ critical behavior of magnetic systems
\cite{gambardella:2002,mukherjeea:2004,dagotto:1996,nguyen:1996,matsumoto:2004}%
, more recently it has also been applied in the study of quantum entanglement,
which plays an essential role in the quantum computation. These applications
can be found in the recent work by Amico \textit{et al}.\cite{Amico:2006}
\ and in the references therein. In particular, in a recent work on quantum
communication in a spin system, Avellino et al.\cite{avellino:2006} have
studied the strong effect of long-range interaction on the fidelity of
transmission of quantum information.

Besides the importance of the long-range interaction in the transmission of
quantum information in spin chains, its presence can induce classical critical
behavior in these systems which is essential for the study of the
classical/quantum crossover.

The one-dimensional XY model is among the models which present this behavior,
provided a uniform long-range interaction is considered. In particular, for
the isotropic one, in the presence of a homogeneous long-range interaction
along the transverse field direction, it can still be solved exactly, and its
solution has been obtained by the authors \cite{goncalves:2001,goncalves:2005}%
. Besides the appearance of classical critical behavior, the most important
features presented by the homogeneous model have been the existence of quantum
bicritical points and, particularly, the existence of first order quantum
transitions. Another relevant result obtained in the study of the homogeneous
case was the exact determination of the classical/quantum crossover for first
and second order phase transitions, and the verification of the scaling
relations proposed Continentino and Ferreira \cite{continentino:2004} for
first order quantum transitions.

As pointed out by Pfleiderer \cite{pfleiderer:2005}, quantum first order
transitions can be driven by different mechanisms, and to look at  this new
quantum critical behavior, in an exactly soluble problem, has been the one of
the motivations to analyze the isotropic model on the inhomogeneous periodic
chain with long-range interaction. Besides this point, the possibility of
having multiple first and second order quantum transitions and multicritical
points of higher order have also been important motivations for considering
the long-range interaction in the inhomogeneous model.

Therefore, in this paper we will extend our previous results
\cite{delima:2006,goncalves:2001,goncalves:2005} by looking at an extended
version of the models previously considered. Although the model presents
classical and quantum critical behavior, we will restrict our analysis to the
quantum one.

In Section 2 we present the exact solution of the model and obtain the
functional of the Helmholtz free energy at arbitrary temperatures. The quantum
critical behavior is discussed in Section 3, and it is determined from the
equation of state at $T=0$. Explicit results for the quantum phase diagrams,
induced magnetization $M^{z}$ and isothermal susceptibility $\chi_{T}^{zz}$
for systems with different sizes of the unit cell are also presented. Finally
in Section 4 we summarize the main results of the paper.

\section{The model and the functional of the free energy}

We consider the isotropic XY model $(s=1/2)$\ on the inhomogeneous periodic
chain with $N$\ cells, $n$\ sites per cell, and lattice parameter $a$, in a
transverse field, with long-range interactions among the spin components in
the $z$ direction. The unit cell of the inhomogeneous lattice is shown in Fig.
1, and the Hamiltonian of the model in its more general form is given by%

\begin{figure}
[ptb]
\begin{center}
\includegraphics[
height=1.4226in,
width=5.1007in
]%
{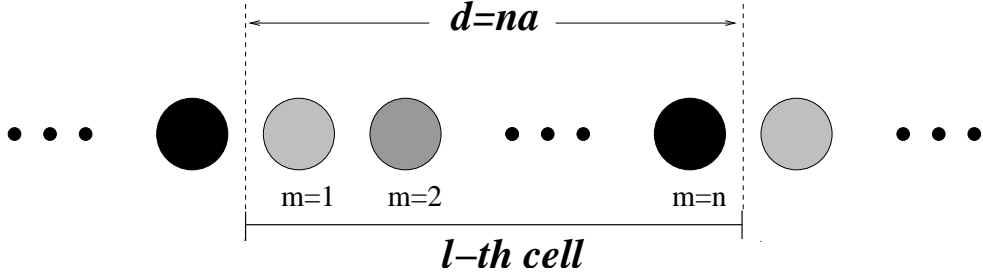}%
\caption{Unit cell of the inhomogeneous chain.}%
\end{center}
\end{figure}
%

\begin{align}
H  &  =-\sum_{l=1}^{N}\left\{  \sum_{m=1}^{n}\mu_{m}hS_{l,m}^{^{z}}+\right.
\sum_{m=1}^{n-1}J_{m}\left[  S_{l,m}^{x}S_{l,m+1}^{x}+S_{l,m}^{y}S_{l,m+1}%
^{y}\right]  +\nonumber\\
&  \left.  +J_{n}S_{l,n}^{x}S_{l+1,1}^{x}+J_{n}S_{l,n}^{y}S_{l+1,1}%
^{y}\right\}  -\frac{J^{\prime}}{N}%
{\displaystyle\sum\limits_{j=1}^{N}}
{\displaystyle\sum\limits_{l=1}^{N}}
\sum_{m,m^{^{\prime}}=1}^{n}S_{j,m}^{z}S_{l,m^{\prime}}^{z},
\label{Hamiltonian}%
\end{align}
where the parameters $J_{m}$\ are the exchange coupling between
nearest-neighbours, $\mu_{m}$\ the magnetic moments, $h$\ the external field,
$J^{\prime}$ the uniform long-range interaction among $z$\ components, and
where we have assumed periodic boundary conditions. If we introduce the ladder
operators%
\begin{equation}
S^{\pm}=S^{x}\pm iS^{y},
\end{equation}
and the Jordan-Wigner transformations
\begin{align}
S_{l,m}^{^{+}}  &  =\exp i\pi\sum_{l^{\prime}=1}^{l-1}\left\{  \sum
_{m^{\prime}=1}^{n}c_{l^{\prime},m^{\prime}}^{\dagger}c_{l^{\prime},m^{\prime
}}+i\pi\sum_{m^{\prime}=1}^{m-1}c_{l,m^{\prime}}^{\dagger}c_{l,m^{\prime}%
}\right\}  c_{l,m}^{\dagger},\\
S_{l,m}^{z}  &  =c_{l,m}^{\dagger}c_{l,m}-\frac{1}{2},
\end{align}

\noindent where $c_{l,m}$\ and $c_{l,m}^{\dagger}$\ are fermion annihilation
and creation operators, we can write the Hamiltonian as\cite{siskens:1974}%
\begin{equation}
H=H^{+}P^{+}+H^{-}P^{-},
\end{equation}
where%
\begin{align}
H^{\pm}  &  =-\sum_{l=1}^{N}\left\{  \sum_{m=1}^{n}\left[  \left(  \mu
_{m}h-J^{\prime}\right)  c_{l,m}^{\dagger}c_{l,m}-\frac{1}{2}\left(  \mu
_{m}h-\frac{J^{\prime}}{2}\right)  \right]  +\sum_{m=1}^{n-1}\frac{J_{m}}%
{2}\left(  c_{l,m}^{\dagger}c_{l,m+1}+c_{l,m+1}^{\dagger}c_{l,m}\right)
\right\}  -\nonumber\\
&  -\sum_{l=1}^{N-1}\frac{J_{n}}{2}\left(  c_{l,n}^{\dagger}c_{l+1,1}%
+c_{l+1,1}^{\dagger}c_{l,n}\right)  \pm\frac{J_{n}}{2}\left(  c_{N,n}%
^{\dagger}c_{1,1}+c_{1,1}^{\dagger}c_{N,n}\right)  -\frac{J^{\prime}}{N}%
{\displaystyle\sum\limits_{j=1}^{N}}
{\displaystyle\sum\limits_{l=1}^{N}}
\sum_{m,m^{^{\prime}}=1}^{n}c_{j,m}^{\dagger}c_{j,m}c_{l,m^{\prime}}^{\dagger
}c_{l,m^{\prime}},
\end{align}
and%
\begin{equation}
P^{\pm}=\frac{I\pm P}{2},
\end{equation}
with $P$ given by
\begin{equation}
P=\exp\left(  i\pi\sum_{l=1}^{N}\sum_{m=1}^{n}c_{l,m}^{\dagger}c_{l,m}\right)
.
\end{equation}

As it is well known \cite{siskens:1974,capel:1977,goncalvestese:1977}, since
the operator $P$ commutes with the Hamiltonian, the eigenstates have definite
parity, and $P^{-}(P^{+})$ corresponds to a projector into a state of odd
(even) parity.

Introducing periodic and anti-periodic boundary conditions on $c^{\prime}s$
for $H^{-}$ and $H^{+}$, respectively, the wave-vectors in the Fourier
transforms
\begin{equation}
c_{l,m}=\frac{1}{\sqrt{N}}\sum_{q}\exp\left(  -iqdl\right)  A_{q,m},
\label{FourierXY}%
\end{equation}

and%

\[
A_{q,m}=\frac{1}{\sqrt{N}}\sum_{l=1}^{N}\exp\left(  iqdl\right)  c_{l,m}%
\]

are given by $q^{-}=\frac{2l\pi}{Nd}$ for periodic condition and $q^{+}%
=\frac{\pi(2l+1)}{Nd}$, for anti-periodic condition \cite{barbosafilho:2001},
with $l=0,\pm1,....,\pm N/2$, and $H^{-}$ and $H^{+}$ can be written in the
form%
\begin{equation}
H^{\pm}=\sum_{q^{\pm}}H_{q^{\pm}},
\end{equation}
where
\begin{align}
H_{q^{\pm}}  &  =-\sum_{m=1}^{n}\left[  \left(  \mu_{m}h-J^{\prime}\right)
A_{q^{\pm},m}^{\dagger}A_{q^{\pm},m}-\frac{1}{2}\left(  \mu_{m}h-\frac
{J^{\prime}}{2}\right)  \right]  -\nonumber\\
&  -\sum_{m=1}^{n-1}\frac{J_{m}}{2}\left[  A_{q^{\pm},m}^{\dagger}A_{q^{\pm
},m+1}+A_{q^{\pm},m+1}^{\dagger}A_{q^{\pm},m}+\gamma_{m}\left(  A_{q^{\pm}%
,m}^{\dagger}A_{q^{\pm},m+1}^{\dagger}+A_{q^{\pm},m+1}A_{q^{\pm},m}\right)
\right]  -\nonumber\\
&  -\frac{J_{n}}{2}\left[  A_{q^{\pm},n}^{\dagger}A_{q^{\pm},1}\exp(-idq^{\pm
})+A_{q^{\pm},1}^{\dagger}A_{q^{\pm},n}\exp(idq^{\pm})\right.  +\nonumber\\
&  +\left.  \gamma_{n}\left(  A_{q^{\pm},n}^{\dagger}A_{q^{\pm},1}^{\dagger
}\exp(-idq^{\pm})+A_{q^{\pm},1}A_{q^{\pm},n}\exp(idq^{\pm})\right)  \right]
-\frac{J^{\prime}}{N}\sum_{m=1}^{n}\left(  A_{q^{\pm},m}^{\dagger}A_{q^{\pm
},m}\right)  ^{2}.
\end{align}
Although $H^{-}$ and $H^{+}$ do not commute, it can be shown that in the
thermodynamic limit all the static properties of the \ system can be obtained
in terms of $H^{-}$ or $H^{+}$
\cite{siskens:1974,capel:1977,goncalvestese:1977}. Therefore, by considering
periodic boundary conditions on $c^{\prime}s,$ we can identify $H\equiv H^{-}$
and $q\equiv q^{-}$, $H_{q}$\ can be written in the form
\begin{align}
H_{q}  &  =-\sum_{m=1}^{n}\left[  \left(  \mu_{m}h-J^{\prime}\right)
A_{q,m}^{\dagger}A_{q,m}-\frac{1}{2}\left(  \mu_{m}h-\frac{J^{\prime}}%
{2}\right)  \right]  -\sum_{m=1}^{n-1}\frac{J_{m}}{2}\left[  A_{q,m}^{\dagger
}A_{q,m+1}+A_{q,m+1}^{\dagger}A_{q,m}\right]  -\nonumber\\
&  -\frac{J_{n}}{2}\left[  A_{q,n}^{\dagger}A_{q,1}\exp(-idq)+A_{q,1}%
^{\dagger}A_{q,n}\exp(idq)\right]  +\text{ }\\
&  +\left.  \gamma_{n}\left(  A_{q,n}^{\dagger}A_{q,1}\exp(-idq)+A_{q,1}%
^{\dagger}A_{q,n}\exp(idq)\right)  \right]  -\frac{J^{\prime}}{N}\left(
\sum_{m=1}^{n}A_{q,m}^{\dagger}A_{q,m}\right)  ^{2}.
\end{align}

The partition function is then given by%
\begin{equation}
Z_{N}=\sqrt{\frac{N}{2\pi}}%
{\displaystyle\prod\limits_{q}}
\left(  Z_{q}\right)  ,
\end{equation}
with%
\begin{align}
&  Z_{q}=\exp\left(  -\frac{\beta}{2}%
{\displaystyle\sum\limits_{m=1}^{n}}
\left(  \mu_{m}h-\frac{J^{\prime}}{2}\right)  \right)  \exp\left\{
{\displaystyle\sum\limits_{m=1}^{n-1}}
\frac{\beta J_{m}}{2}\left[  A_{q,m}^{\dagger}A_{q,m+1}+A_{q,m+1}^{\dagger
}A_{q,m}\right]  +\right. \nonumber\\
&  +\frac{\beta J_{n}}{2}\left[  A_{q,n}^{\dagger}A_{q,1}\exp(-idq)+A_{q,1}%
^{\dagger}A_{q,n}\exp(idq)\right]  +%
{\displaystyle\sum\limits_{m=1}^{n}}
\beta\left(  \mu_{m}h-J^{\prime}\right)  A_{q,m}^{\dagger}A_{q,m}\nonumber\\
&  \left.  +\left(  \sqrt{\frac{\beta J^{\prime}}{N}}\sum_{m=1}^{n}%
A_{q,m}^{\dagger}A_{q,m}\right)  ^{2}\right\}  .
\end{align}
Since the long-range interaction term commutes with the Hamiltonian, we can
introduce the gaussian transformation
\begin{equation}
\exp\left(  b^{2}\right)  =\frac{1}{\sqrt{2\pi}}%
{\displaystyle\int\limits_{-\infty}^{\infty}}
\exp\left(  -\frac{x^{2}}{2}+\sqrt{2}bx\right)  dx,
\end{equation}
in the previous expression, so that the partition function can be rewritten in
an integral representation as
\[
Z_{q}=\exp\left(  -\frac{\beta}{2}%
{\displaystyle\sum\limits_{m=1}^{n}}
\left(  \mu_{m}h-\frac{J^{\prime}}{2}\right)  \right)  \int_{-\infty}^{\infty
}\exp\left(  -\frac{\overline{x}^{2}}{2}\right)  Tr\left[  \exp\left(
-\beta\widetilde{H_{q}}(\overline{x})\right)  \right]  d\overline{x},
\]
where $\overline{x}=x/\sqrt{nN,}$ and the effective Hamiltonian $\widetilde
{H}_{q}(\overline{x})$ is given by%

\begin{align}
\widetilde{H}_{q}(\overline{x})  &  =-%
{\displaystyle\sum\limits_{m=1}^{n-1}}
\frac{J_{m}}{2}\left[  A_{q,m}^{\dagger}A_{q,m+1}+A_{q,m+1}^{\dagger}%
A_{q,m}\right]  -\frac{J_{n}}{2}\left[  A_{q,n}^{\dagger}A_{q,1}%
\exp(-idq)+A_{q,1}^{\dagger}A_{q,n}\exp(idq)\right]  -\nonumber\\
&  -%
{\displaystyle\sum\limits_{m=1}^{n}}
\left(  \mu_{m}h-J^{\prime}+\sqrt{\frac{2J^{\prime}}{\beta}}\overline
{x}\right)  A_{q,m}^{\dagger}A_{q,m}.
\end{align}

By introducing the canonical transformations%
\begin{equation}
A_{q,m}=\sum_{k=1}^{n}u_{q,km}\xi_{q,k},\text{ \ \ \ \ }A_{q,m}^{\dag}%
=\sum_{k=1}^{n}u_{q,km}^{\ast}\xi_{q,k}^{\dag}, \label{transformationI}%
\end{equation}
and by imposing the condition
\begin{equation}
\lbrack\xi_{q,k},\widetilde{H}_{q}(\overline{x})]=\varepsilon_{q,k}\xi_{q,k},
\label{comutator}%
\end{equation}
this leads, for the coefficients $u_{q,km},$\ to the equation
\begin{equation}
\mathbf{A}_{q}%
\begin{pmatrix}
u_{q,k1}\\
u_{q,k2}\\
\vdots\\
u_{q,kn}%
\end{pmatrix}
=\varepsilon_{q,k}%
\begin{pmatrix}
u_{q,k1}\\
u_{q,k2}\\
\vdots\\
u_{q,kn}%
\end{pmatrix}
, \label{eigenequation}%
\end{equation}
where $\mathbf{A}_{q}$\ is given by%
\begin{equation}
\mathbf{A}_{q}\equiv-%
\begin{pmatrix}
\widetilde{h}_{1} & \frac{J_{1}}{2} & 0 & \cdots & 0 & \frac{J_{n}}{2}%
\exp\left(  -iqd\right) \\
\frac{J_{1}}{2} & \widetilde{h}_{2} & \frac{J_{2}}{2} &  &  & 0\\
0 & \frac{J_{2}}{2} & \widetilde{h}_{3} & \frac{J_{3}}{2} &  & \vdots\\
\vdots &  & \frac{J_{3}}{2} & \ddots & \ddots & 0\\
0 &  &  & \ddots & \widetilde{h}_{n-1} & \frac{J_{n-1}}{2}\\
\frac{J_{n}}{2}\exp\left(  iqd\right)  & 0 & \cdots & 0 & \frac{J_{n-1}}{2} &
\widetilde{h}_{n}%
\end{pmatrix}
, \label{Aq}%
\end{equation}
and the $u^{\prime}s$\ satisfy the orthogonality relations,%
\begin{gather}
\sum\limits_{m=1}^{n}u_{q,km}u_{q,k^{\prime}m}^{\ast}=\delta_{kk^{\prime}%
},\label{Ortogonality1}\\
\sum\limits_{k=1}^{n}u_{q,km}u_{q,km^{\prime}}^{\ast}=\delta_{mm^{\prime}},
\label{Ortogonality2}%
\end{gather}
and $\widetilde{h}_{j}=\mu_{m}h-J^{\prime}+\sqrt{\frac{2J^{\prime}}{\beta}%
}\overline{x}.$ Therefore the effective Hamiltonian can be written in the
diagonal form
\begin{equation}
\widetilde{H}_{q}(\overline{x})=\sum_{k}\widetilde{\varepsilon}_{q,k}%
\eta_{q,k}^{\dagger}\eta_{q,k}, \label{freefermions}%
\end{equation}
where the spectrum $\widetilde{\varepsilon}_{q}$\ of $H_{q}$\ is determined
from the determinantal equation%
\begin{equation}
det(\mathbf{A}_{q}-\widetilde{\varepsilon}_{q}\mathbf{I})=0,
\label{dispersion}%
\end{equation}
and the operators $\eta^{\prime}s$\ are given in terms of $A^{\prime}s$\ and
$c^{\prime}s$ by the expression
\begin{equation}
\eta_{q,k}=\sum\limits_{m=1}^{n}u_{q,km}^{\ast}A_{q,m}=\frac{1}{\sqrt{N}}%
\sum\limits_{l=1}^{N}\sum\limits_{m=1}^{n}\exp\left(  iqdl\right)
u_{q,km}^{\ast}c_{l,m},
\end{equation}
which have been obtained by using eqs.(\ref{FourierXY}) and
(\ref{Ortogonality1}). From this result the partition function can be written
in the form

\bigskip%
\begin{align}
Z_{N}  &  =\sqrt{\frac{N}{2\pi}}\exp\left[  -\frac{\beta N}{2}%
{\displaystyle\sum\limits_{m=1}^{n}}
\left(  \mu_{m}h-\frac{J^{\prime}}{2}\right)  \right]  \int_{-\infty}^{\infty
}\exp\left(  -\frac{nN\overline{x}^{2}}{2}\right)  \times\\
&  \times\exp\left(  \sum_{q,k}\ln\left(  1+\exp(-\beta\widetilde{\varepsilon
}_{q,k}\right)  )\right)  d\overline{x}.
\end{align}

In the thermodynamic limit, the partition function can be evaluated by
\ Laplace%
\'{}%
s method by imposing the condition that $g^{\prime}(\overline{x}_{0})=0,$
where $g(\overline{x})$ is given by%
\[
g(\overline{x})=-\frac{\overline{x}^{2}}{2}+\frac{1}{nN}\sum_{q,k}\left[
\ln(1+\exp(-\beta\widetilde{\varepsilon}_{q,k}))\right]  ,
\]
and \ $\overline{x}_{0}$ is equal to%
\[
\overline{x}_{0}=\frac{1}{nN}\sum_{q,k}\frac{\sqrt{2\beta J^{\prime}}}%
{1+\exp(-\beta\widetilde{\varepsilon}_{q,k})},
\]
which can be written in terms of the average induced magnetization $M^{z}$,
defined as%
\begin{equation}
M^{z}=\frac{1}{nN}\sum_{l,m}\mu_{m}\langle S_{l,m}^{z}\rangle=\frac{1}{nN}%
\sum_{q,k,m}\mu_{m}u_{q,km}^{\ast}u_{q,km}\langle\eta_{q,k}^{\dag}\eta
_{q,k}\rangle-\frac{1}{2}, \label{Magnetization}%
\end{equation}
in the form%
\[
\frac{\overline{x}_{0}}{\sqrt{2\beta J^{\prime}}}=M^{z}+\frac{1}{2}.
\]

Finally, we can obtain from the previous results the functional of the
Helmholtz free energy per lattice site which is given by%

\begin{equation}
f=\frac{h}{2}-\frac{k_{B}T}{nN}\sum_{q,k}\ln\left[  1+\exp(-\beta
\widetilde{\varepsilon}_{q,k})\right]  +J^{\prime}M^{z}(M^{z}+1).
\label{Functional}%
\end{equation}

The equation of state is obtained numerically from this functional by imposing
the conditions

\begin{equation}
\frac{\partial f}{\partial M^{z}}=0,\ \ \ \text{with }\frac{\partial^{2}%
f}{\partial M^{z2}}>0. \label{Equilibrium conditions}%
\end{equation}
The numerical solution is more easily obtained for $n\leq4,$ where there are
analytical solutions for eqs.(\ref{eigenequation}) and (\ref{dispersion})
\cite{delima:2006}$,$ and in the cases where we have uniform magnetic moments,
namely, $\mu_{m}\equiv\mu.$ In this situation, \ the term $-\sum_{l,m}^{n}\mu
hS_{l,m}^{^{z}}$\ commutes with the Hamiltonian, and consequently the effect
of the field is to shift the spectrum. This means that the solution of
eq.(\ref{dispersion}) can be written as%

\begin{equation}
\widetilde{\varepsilon}_{q,k}=\widetilde{\varepsilon}_{q,k}^{0}-\mu
h+J^{\prime}-\sqrt{\frac{2J^{\prime}}{\beta}}\overline{x_{0}}=\widetilde
{\varepsilon}_{q,k}^{0}-\mu h-2J^{\prime}M^{z}, \label{Energy excitations}%
\end{equation}
where $\widetilde{\varepsilon}_{q,k}^{0}$ is the energy of the excitations of
the model for zero transverse field and long-range interaction, and we
identify the expression $\mu h+2J^{\prime}M^{z}$ as an effective field
$h_{eff}$ \cite{goncalves:2005}.

\section{The quantum critical behavior}

In the limit $T\rightarrow0,$ the functional of the Helmholtz free energy per
lattice site, eq.(\ref{Functional}), can be explicitly written as
\begin{equation}
f=\frac{F_{N}}{N}=\frac{h}{2}+J^{\prime}M^{z}(M^{z}+1)-\frac{1}{\pi n}%
{\displaystyle\sum\limits_{k=1}^{n}}
{\displaystyle\int\limits_{0}^{\overline{q}_{k}}}
\varepsilon_{q,k}dq_{k},
\end{equation}
where $\overline{q}_{k}$ is obtained by imposing the condition $\widetilde
{\varepsilon}_{q,k}=0$ and we have considered the lattice spacing $a=1$ and
uniform magnetic moments $\mu=1.$

Under these conditions the average induced magnetization, $M^{z}$, given in
eq.(\ref{Magnetization}), can also be written in the form
\begin{equation}
M^{z}=\frac{1}{2\pi n}%
{\displaystyle\sum\limits_{k=1}^{n}}
{\displaystyle\int\limits_{0}^{\overline{q}_{k}}}
sign(\varepsilon_{q,k})dq_{k}.
\end{equation}

The quantum phase diagram is determined from the equation of state, which is
obtained numerically from the previous equations by considering the conditions
shown in eq.(\ref{Equilibrium conditions}), and in particular we can determine
the magnetization as a function of the transverse field. \ As it is well
known, the second order phase transitions are determined by imposing the limit
$M^{z}\rightarrow0$ and it can be shown numerically, for arbitrary unit cell
sizes, that the critical field varies linearly with the long-range range
interaction $J^{\prime}$ and that these phase transitions, as in the
homogenous model \cite{goncalves:2005}, occur for $J^{\prime}<0$ only.

The first order phase transitions are determined by imposing the additional condition

\begin{equation}
f(M^{z})=f\left(  M_{p}^{z}\right)  ,
\end{equation}
where $M_{p}^{z}$ are the magnetization plateaus, which are identical to the
ones in the model without the long-range interaction \cite{delima:2006}. This
is due to the fact that, for uniform magnetic moments, the long-range
interaction does preserve the azimutal symmetry of the model and consequently
the magnetization plateaus satisfy the quantization condition
\cite{oshikawa:1997}%

\begin{equation}
n\left(  \frac{\mu}{2}-M_{p}^{z}\right)  =\mu\times\operatorname{integer},
\end{equation}
\bigskip

\noindent which does only depend on the symmetry of the Hamiltonian.

As in the homogenous case \cite{goncalves:2005}, the first order phase
transitions occur for $J^{\prime}>0$ only, and in this case, besides the
bicritical points where the second order lines meet the first order ones
\cite{goncalves:2001}, there are triple points which correspond to the point
where three first order lines meet.

From the magnetization as a function of the field we can obtain the isothermal
susceptibility, $\chi_{T}^{zz},$ which is given by
\begin{equation}
\chi_{T}^{zz}\equiv\frac{1}{n}\frac{\partial M^{z}}{\partial h}.
\label{susceptibilityisothermal}%
\end{equation}

The main results are shown in Figs. 2 \ to 11. In Fig. 2 we present the phase
diagram for $n=2,$ and as expected there are two second order lines for
$J^{\prime}<0$ which end up at bicritical points for $J^{\prime}=0$. For
$J^{\prime}>0$, there are two first order lines which collapse into a single
one at a triple point. In Fig. 3(a) we present the magnetization as function
of the field, and as it can be seen the different critical behaviors are
explicitly shown depending on the sign of the long-range interaction.\ It
should be noted that as we approach the triple point the two first order phase
transitions collapse into a single one, which corresponds in this case to a
jump in the induced magnetization from zero to one half.

As in the homogeneous model\cite{goncalves:2005}, there is a universal curve
to which all magnetization data collapse, independently of the order of the
transition. This is presented in Fig. 3(b), where we show the magnetization as
a function of the effective field $h_{eff}=h+2J^{\prime}M^{z}.$%

\begin{figure}
[ptb]
\begin{center}
\includegraphics[
height=6.8554in,
width=4.1286in
]%
{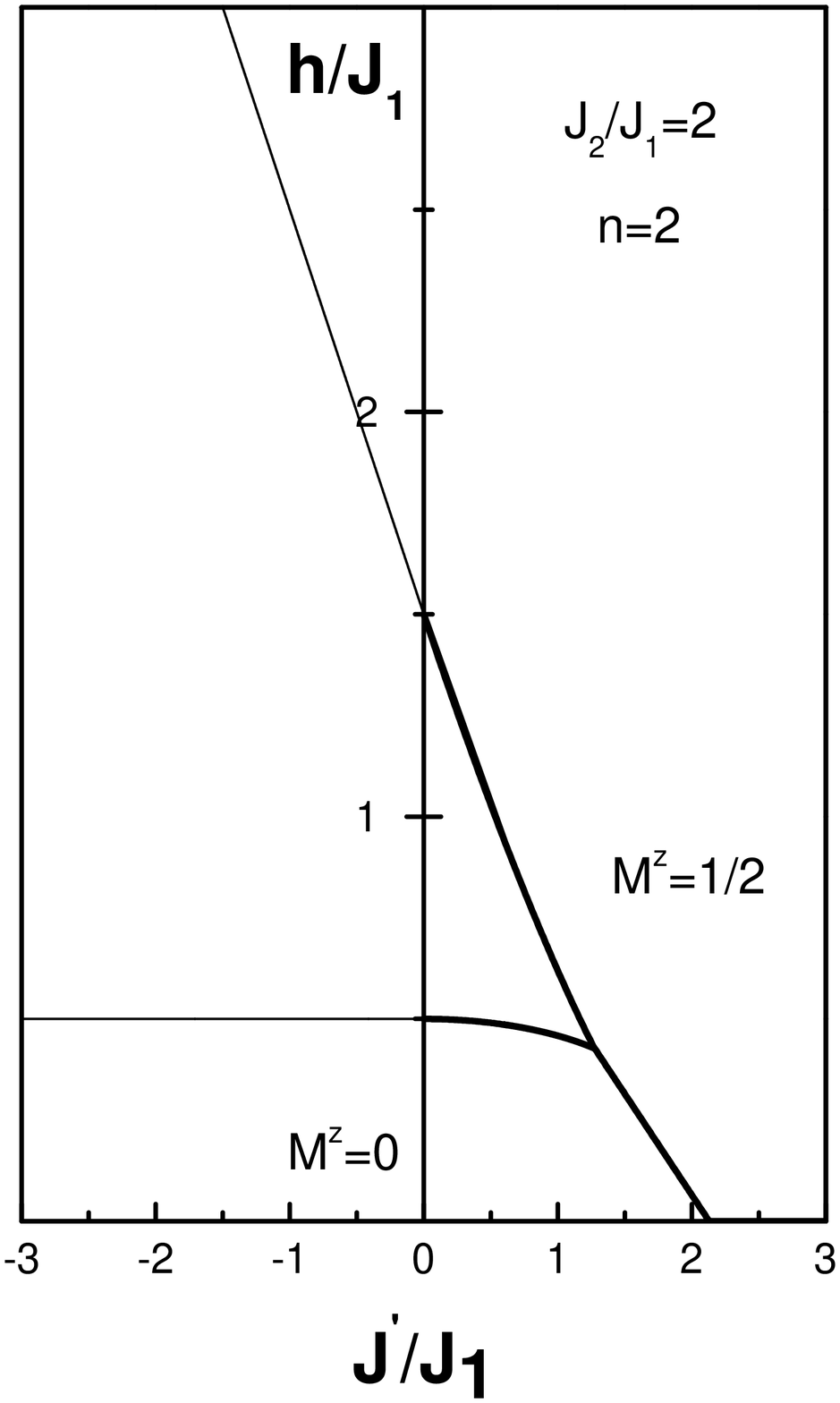}%
\caption{Phase diagram for the quantum transitions as a function of the
strength of the\ long-range interaction $J^{\prime}/J_{1},$ for $n=2$ and
$J_{1}=1,$ $J_{2}=2.$ For $J^{\prime}/J_{1}>0,$ the critical lines identify
the first-order phase transitions and, for $J^{\prime}/J_{1}\leq0,$ the second
order phase transitions. }%
\end{center}
\end{figure}

In Fig.4 we present the quantum phase diagram for $n=3.$ As in the previous
case, for $J^{\prime}<0$ we have second order phase transitions and for
$J^{\prime}>0$ we have first order phase transitions. There are three second
order lines and two first order lines meet at a unique triple point. The
magnetization is presented in Fig.5(a)\ for different values of $J^{\prime}$,
which characterize the different behaviors, and in Fig 5(b) we present the
collapse of the magnetization when plotted as a function of the effective
field $h_{eff}.$

It should be noted that for a different set of parameters we could have two
triple points, and in this case we will have a single first order line beyond
the critical value $J^{\prime},$ which is associated to the second triple
point. This situation can be seen in the phase diagram shown in Fig 6, for
$n=4$, where it is explicitly shown that we can have three triple points.
Since we have no suppression of a phase transition, there are four second
order transition lines. As it has been shown for $J^{\prime}=0$
\cite{delima:2006}, an adequate choice of the exchange parameters can suppress
a phase transition. For $n=4$, this condition corresponds to $J_{1}J_{3}%
=J_{2}J_{4}$ and the phase diagram for this case is presented in Fig. 7. As it
can be verified in this figure, we have three second order lines, instead of
four, and just one triple point.

The magnetization associated to the parameters defined in Fig. 6 is presented
in Fig. 8. As in the previous cases, in Fig 8(a) we have the magnetization as
a function of the field for different values of $J^{\prime}$, and in Fig.8(b),
the universal curve for the magnetization as a function of the effective field
$h_{eff}.$ In Fig. 9, we present the phase diagram for $n=5.$ Since $n$ is
odd, there is no suppression of any transition for $J^{\prime}=0$
\cite{delima:2006}, and, as expected, there are five second order transition
lines. The associated magnetization as a function of the field is presented in
Fig 10(a) and the magnetization universal curve is shown in Fig. 10(b).

Finally, in Fig. 11, we present the isothermal susceptibility, $\chi_{T}%
^{zz},$ for $n=3$ and different values of $J^{\prime}.$ As in the homogeneous
model, the isothermal susceptibility, at the second order phase transitions,
diverges for $J^{\prime}=0$ only, and its multiple phase transitions have the
same critical exponents of the homogeneous model and consequently belong to
same universality class.

\section{ Conclusions}

In this work we have considered the isotropic XY model with an uniform
long-range interaction along the $z$ direction, in a periodic inhomogenous
lattice with $N$ cells and $n$ sites per cell. The exact solution of the model
was formally obtained at arbitrary temperatures and distribution of \ magnetic
moments and exchange constants. Explicit equations have been obtained for the
functional of the Helmholtz free energy from which the equation of state can
be determined numerically.

The analysis of the critical behavior has been restricted to the quantum phase
transitions and we have shown, as in the homogeneous model, that the system
presents first order phase transitions when the long-range interaction is
ferromagnetic, and second order phase transitions when the long-range
interaction is antiferromagnetic. The model also presents multiple first and
second order phase transitions and the number of critical lines is equal to
$n,$ for $n$ odd, and less than $n,$ for $n$ even, provided the exchange
constants in the $xy$ plane satisfy a special relation.

The second order critical lines meet the first order ones at $J^{\prime}=0,$
which are bicritical points and there are multiple triple points, where three
first order critical lines meet, depending on the unit cell size. The critical
exponents have been obtained numerically, and it has been shown, as expected,
that the model belongs to the same universality class of the homogeneous one.

Finally, we would like to point out that is of paramount importance the
presence of multiple first order quantum transitions and triple points, which
we have shown exactly to exist in the model, since we believe that this
behavior is related to the different mechanisms from which the first order
phase transitions are driven. These mecanisms have been thoroughly discusssed
by Pfleiderer\cite{pfleiderer:2005}, by analyzing quantum critical behavior
obtained for different materials.%

\begin{figure}
[ptb]
\begin{center}
\includegraphics[
height=6.8528in,
width=4.2237in
]%
{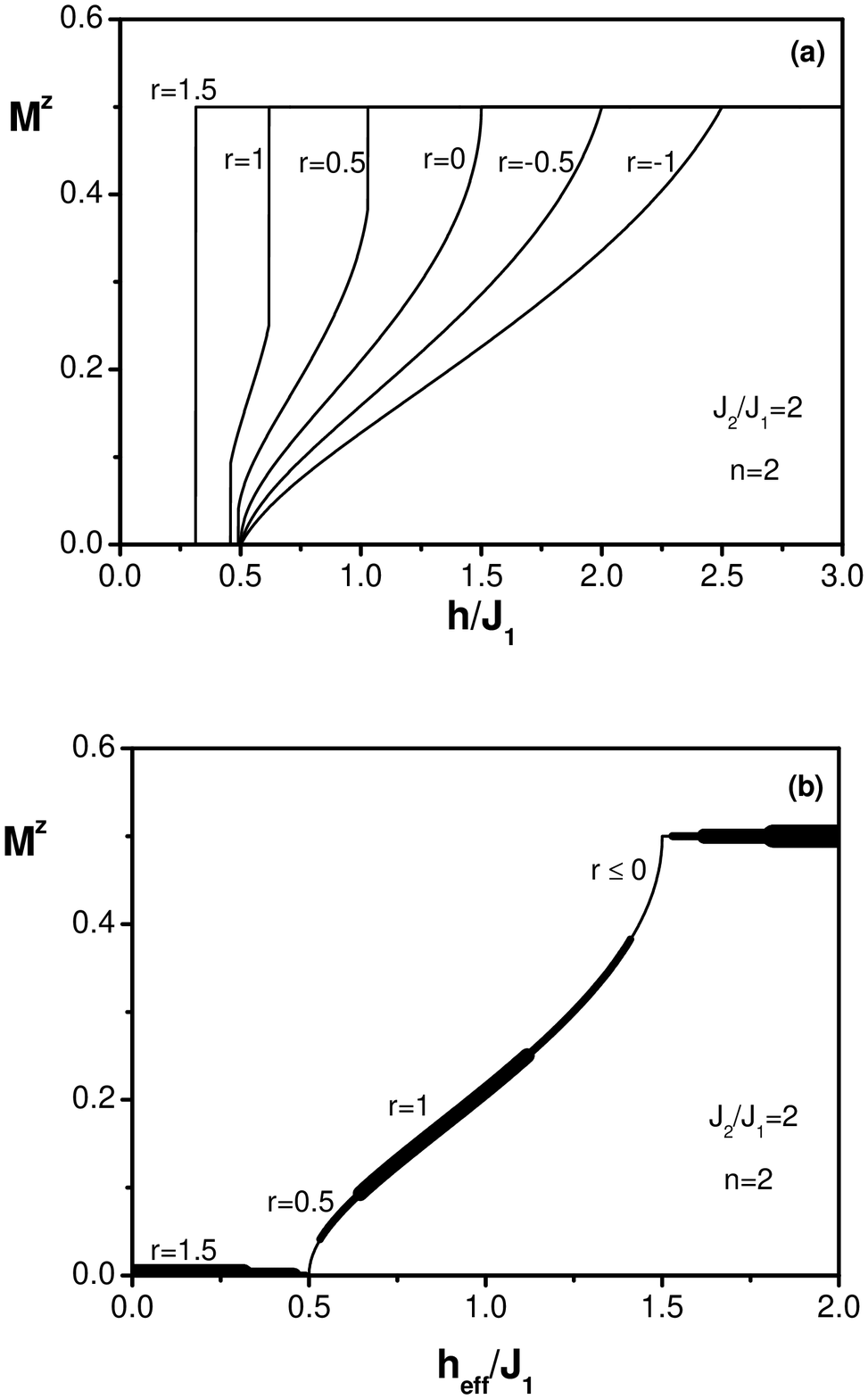}%
\caption{(a) Magnetization as a function of $h/J_{1}$, and (b) universal curve
for the magnetization as a function of the effective field $h_{eff}/J_{1}$
$(h_{eff}=h+2J^{\prime}M^{z}),$ at $T=0,$ for different values of
$r$($r=J^{\prime}/J_{1})$, in the regions where the system undergoes first
($r>0)$ and second-order ($r\leqslant0)$ quantum transitions, for $n=2$ and
$J_{1}=1,$ $J_{2}=2.$}%
\end{center}
\end{figure}

%

\begin{figure}
[ptb]
\begin{center}
\includegraphics[
height=6.8571in,
width=4.369in
]%
{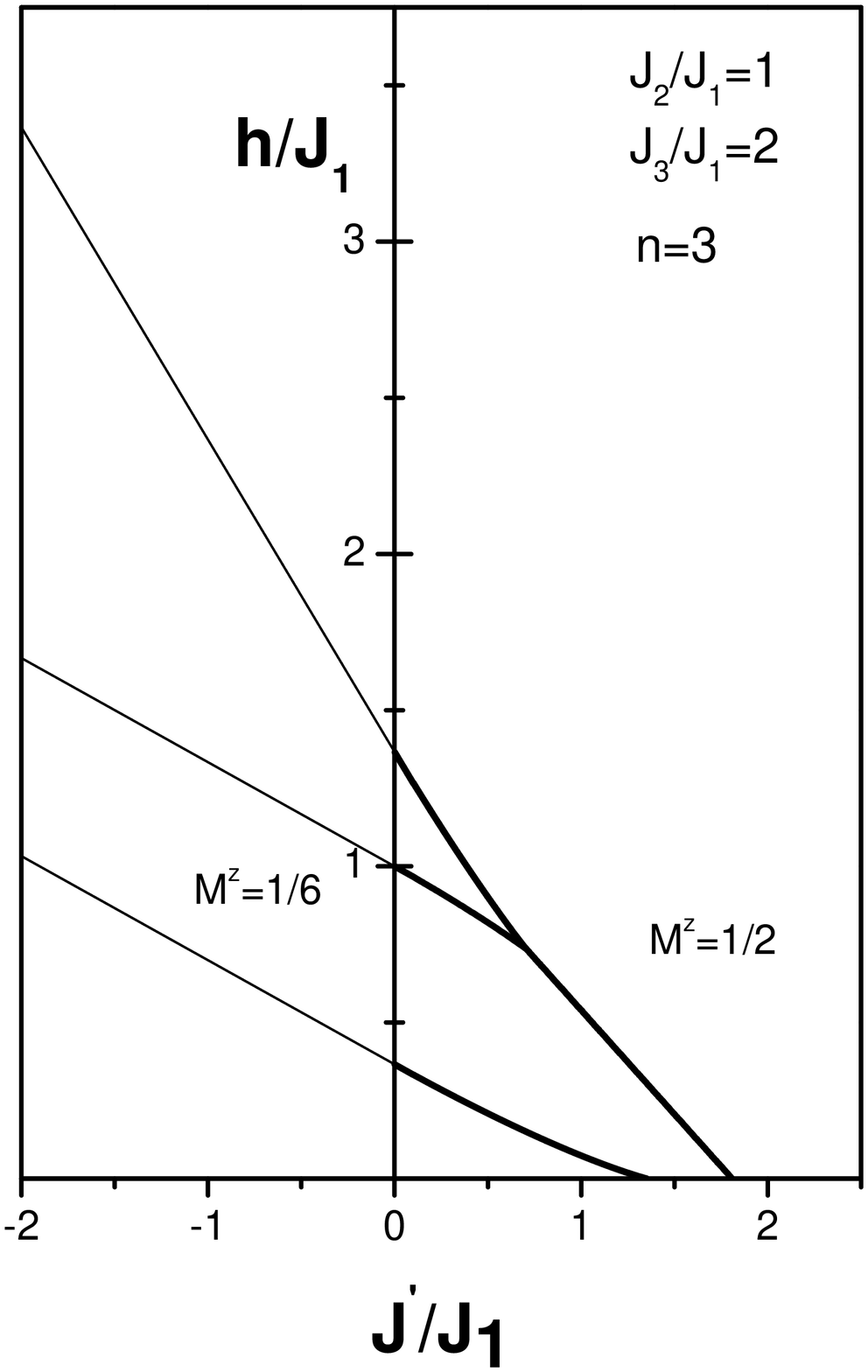}%
\caption{Phase diagram for the quantum transitions as a function of the
strength of the\ long-range interaction $J^{\prime}/J_{1},$ for $n=3$ and
$J_{1}=1,$ $J_{2}=1,J_{3}=2.$ For $J^{\prime}/J_{1}>0,$ the critical lines
identify the first-order phase transitions and, for $J^{\prime}/J_{1}\leq0,$
the second order phase transitions. }%
\end{center}
\end{figure}
%

\begin{figure}
[ptb]
\begin{center}
\includegraphics[
height=6.8571in,
width=3.8735in
]%
{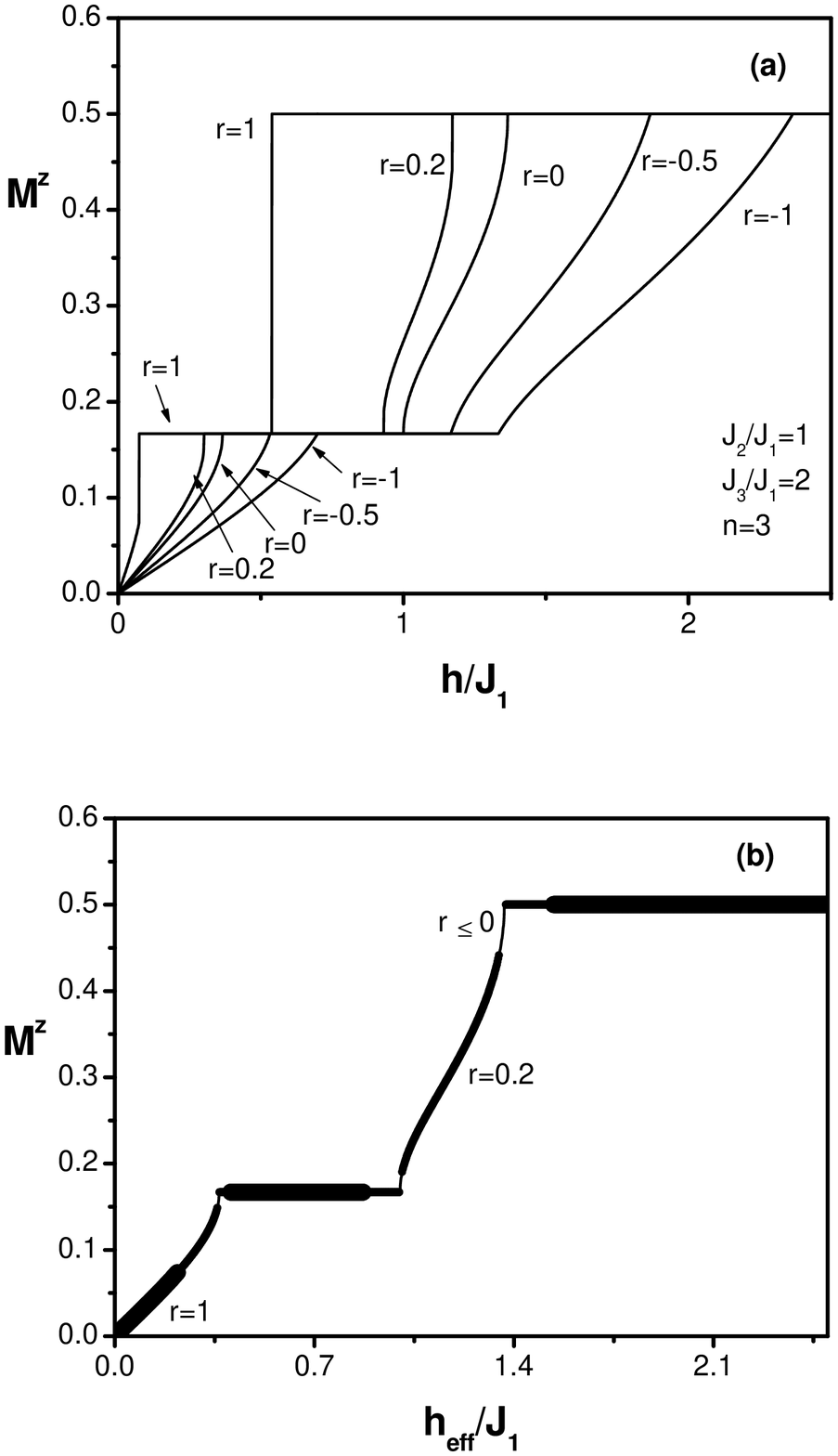}%
\caption{(a) Magnetization as a function of $h/J_{1}$, and (b) universal curve
for the magnetization as a function of the effective field $h_{eff}/J_{1}$
$(h_{eff}=h+2J^{\prime}M^{z}),$ at $T=0,$ for different values of
$r$($r=J^{\prime}/J_{1})$, in the regions where the system undergoes first
($r>0)$ and second-order ($r\leqslant0)$ quantum transitions, for $n=3$ and
$J_{1}=1,$ $J_{2}=1,J_{3}=2.$}%
\end{center}
\end{figure}

%

\begin{figure}
[ptb]
\begin{center}
\includegraphics[
height=6.858in,
width=4.1018in
]%
{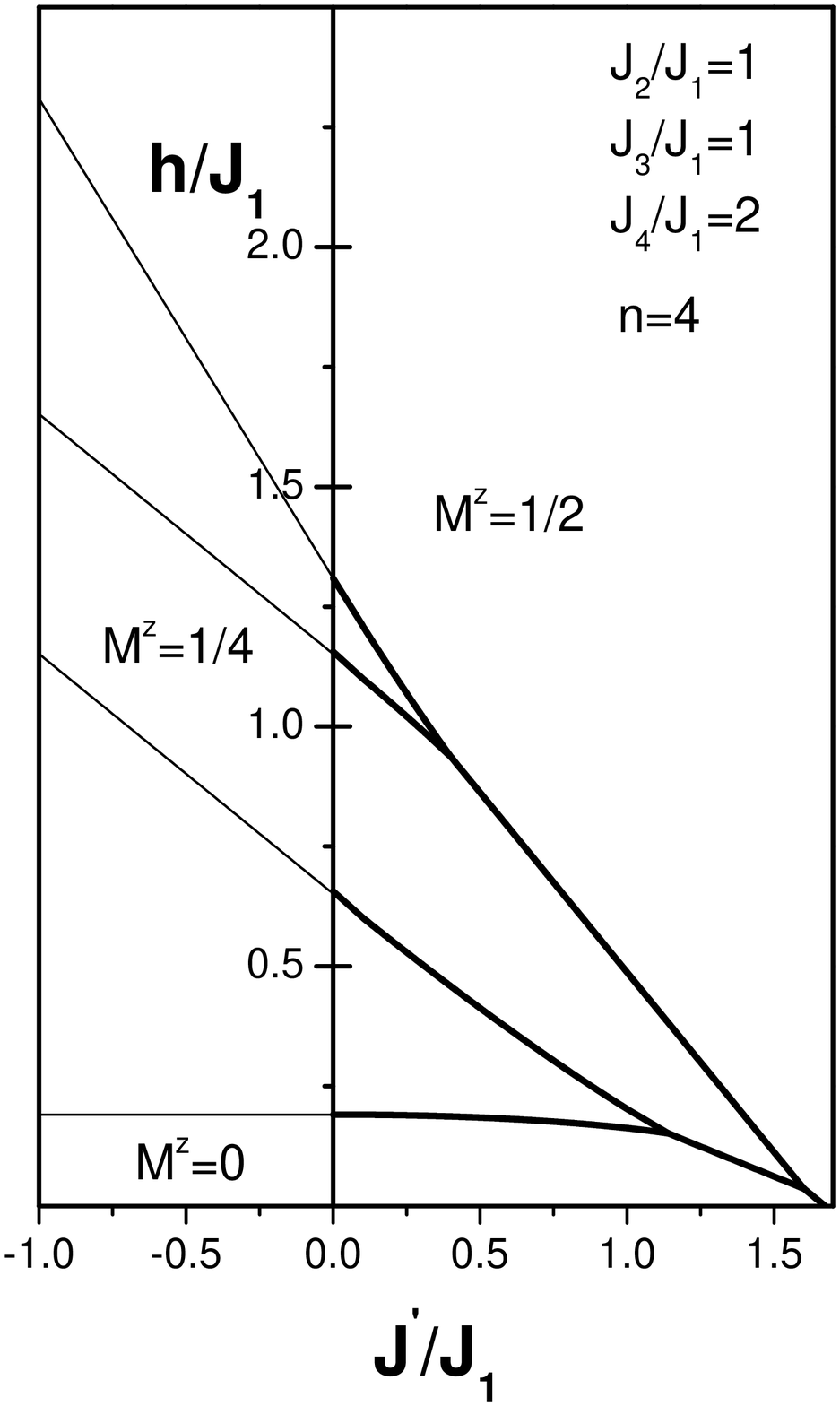}%
\caption{Phase diagram for the quantum transitions as a function of the
strength of the\ long-range interaction $J^{\prime}/J_{1},$ for $n=4$ and
$J_{1}=1,$ $J_{2}=1,J_{3}=1,J_{4}=2.$ For $J^{\prime}/J_{1}>0,$ the critical
lines identify the first-order phase transitions and, for $J^{\prime}%
/J_{1}\leq0,$ the second order phase transitions. }%
\end{center}
\end{figure}

%

\begin{figure}
[ptb]
\begin{center}
\includegraphics[
height=6.858in,
width=4.2263in
]%
{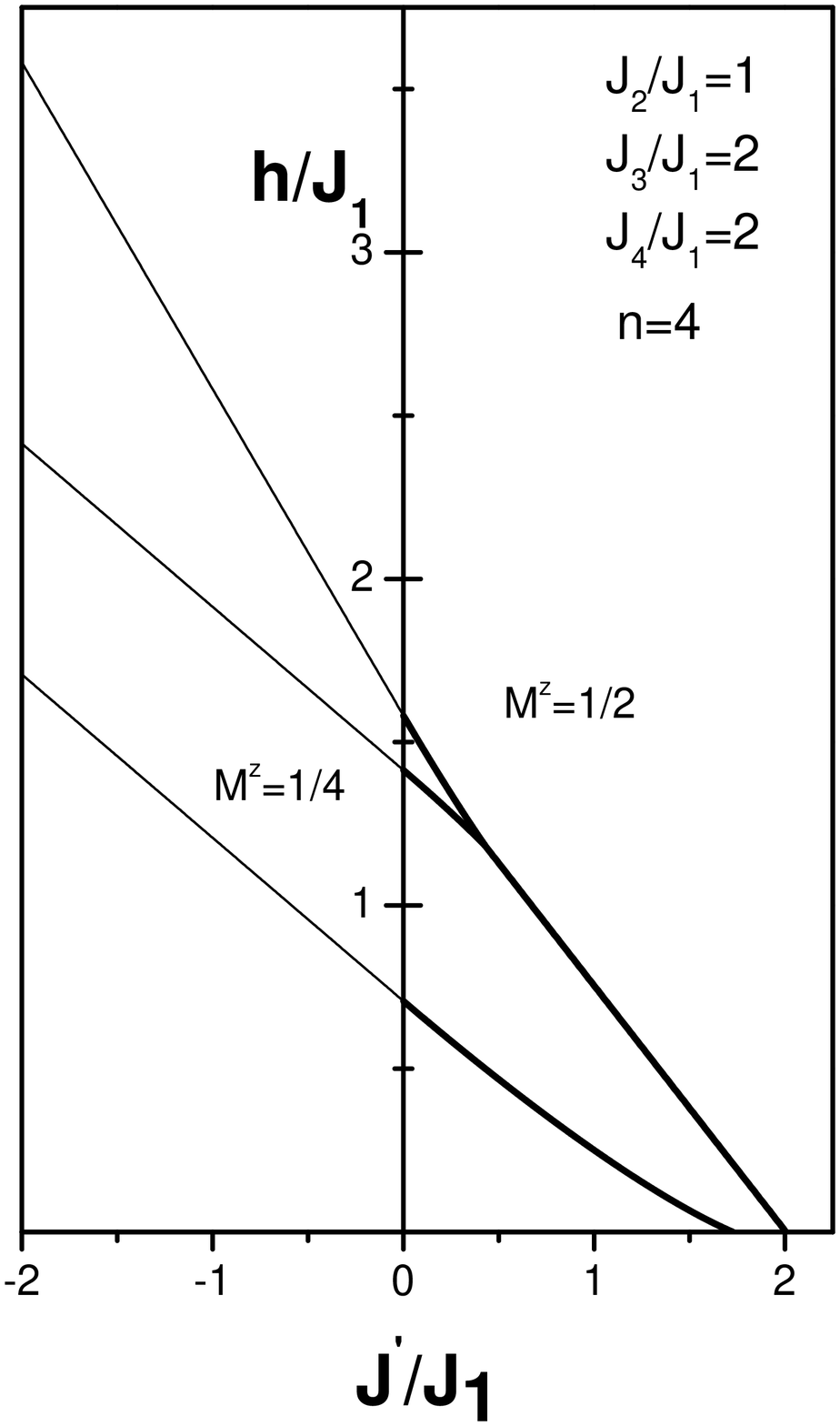}%
\caption{Phase diagram for the quantum transitions as a function of the
strength of the\ long-range interaction $J^{\prime}/J_{1},$ for $n=4$ and
$J_{1}=1,$ $J_{2}=1,J_{3}=2,J_{4}=2.$ For $J^{\prime}/J_{1}>0,$ the critical
lines identify the first-order phase transitions and, for $J^{\prime}%
/J_{1}\leq0,$ the second order phase transitions. }%
\end{center}
\end{figure}
%

\begin{figure}
[ptb]
\begin{center}
\includegraphics[
height=6.8571in,
width=4.8577in
]%
{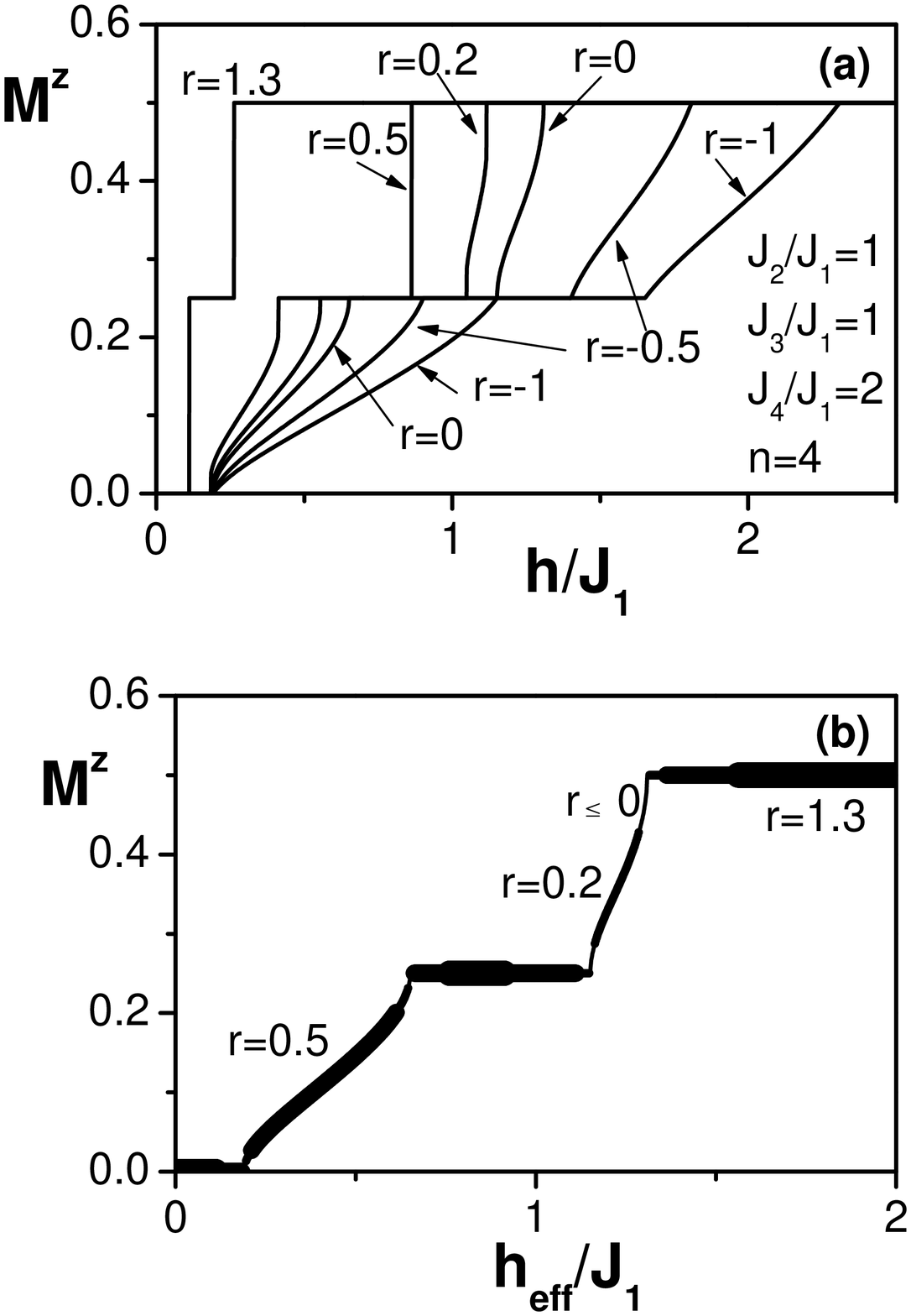}%
\caption{(a) Magnetization as a function of $h/J_{1}$, and (b) universal curve
for the magnetization as a function of the effective field $h_{eff}/J_{1}$
$(h_{eff}=h+2J^{\prime}M^{z}),$ at $T=0,$ for different values of
$r$($r=J^{\prime}/J_{1})$, in the regions where the system undergoes first
($r>0)$ and second-order ($r\leqslant0)$ quantum transitions, for $n=4$ and
$J_{1}=1,$ $J_{2}=1,J_{3}=1,J_{4}=2.$}%
\end{center}
\end{figure}

%

\begin{figure}
[ptb]
\begin{center}
\includegraphics[
height=6.858in,
width=4.2886in
]%
{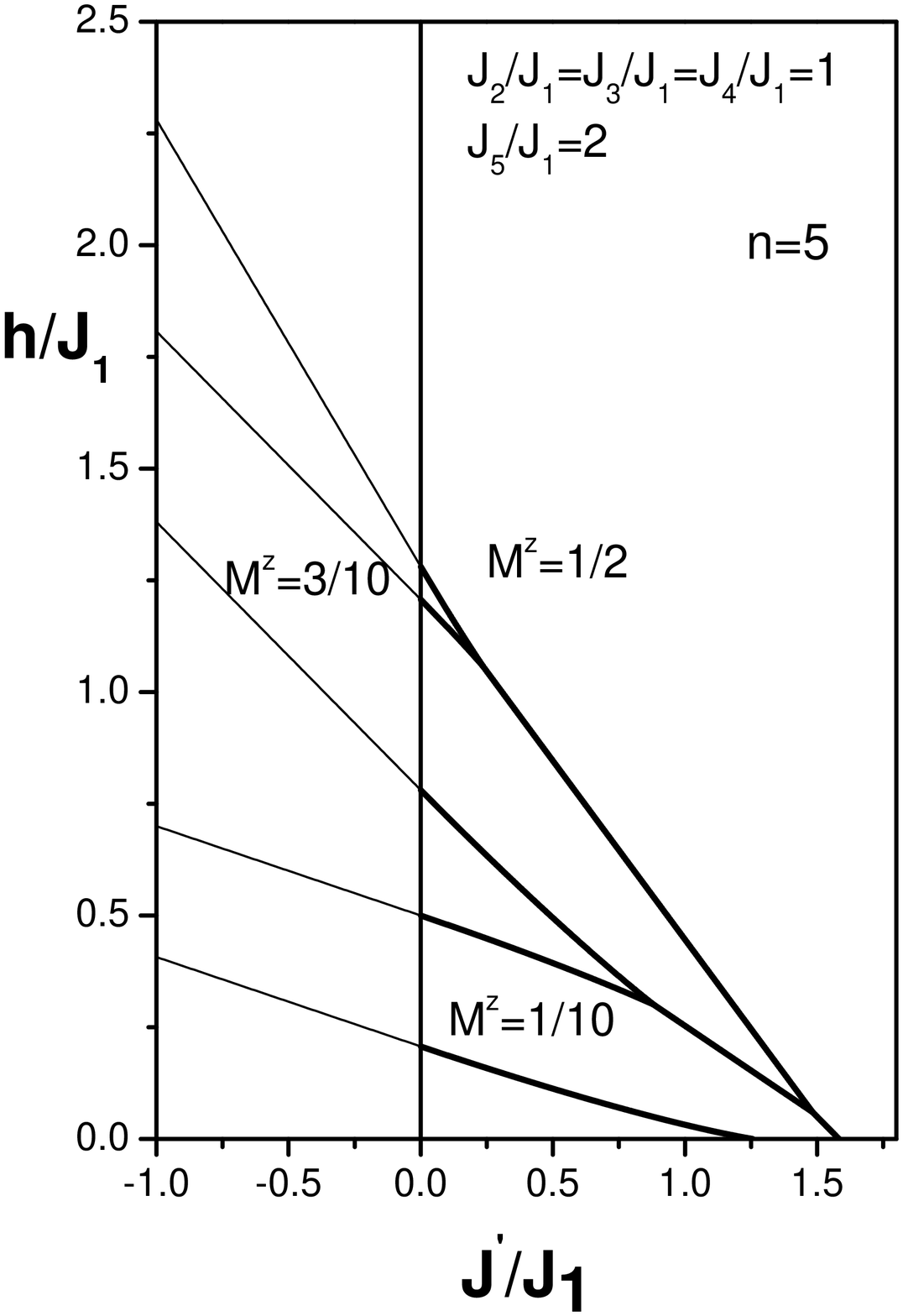}%
\caption{Phase diagram for the quantum transitions as a function of the
strength of the\ long-range interaction $J^{\prime}/J_{1},$ for $n=5$ and
$J_{1}=1,$ $J_{2}=1,J_{3}=1,J_{4}=1,J_{5}=2.$ For $J^{\prime}/J_{1}>0,$ the
critical lines identify the first-order phase transitions and, for $J^{\prime
}/J_{1}\leq0,$ the second order phase transitions. }%
\end{center}
\end{figure}

%

\begin{figure}
[ptb]
\begin{center}
\includegraphics[
height=6.8545in,
width=5.3324in
]%
{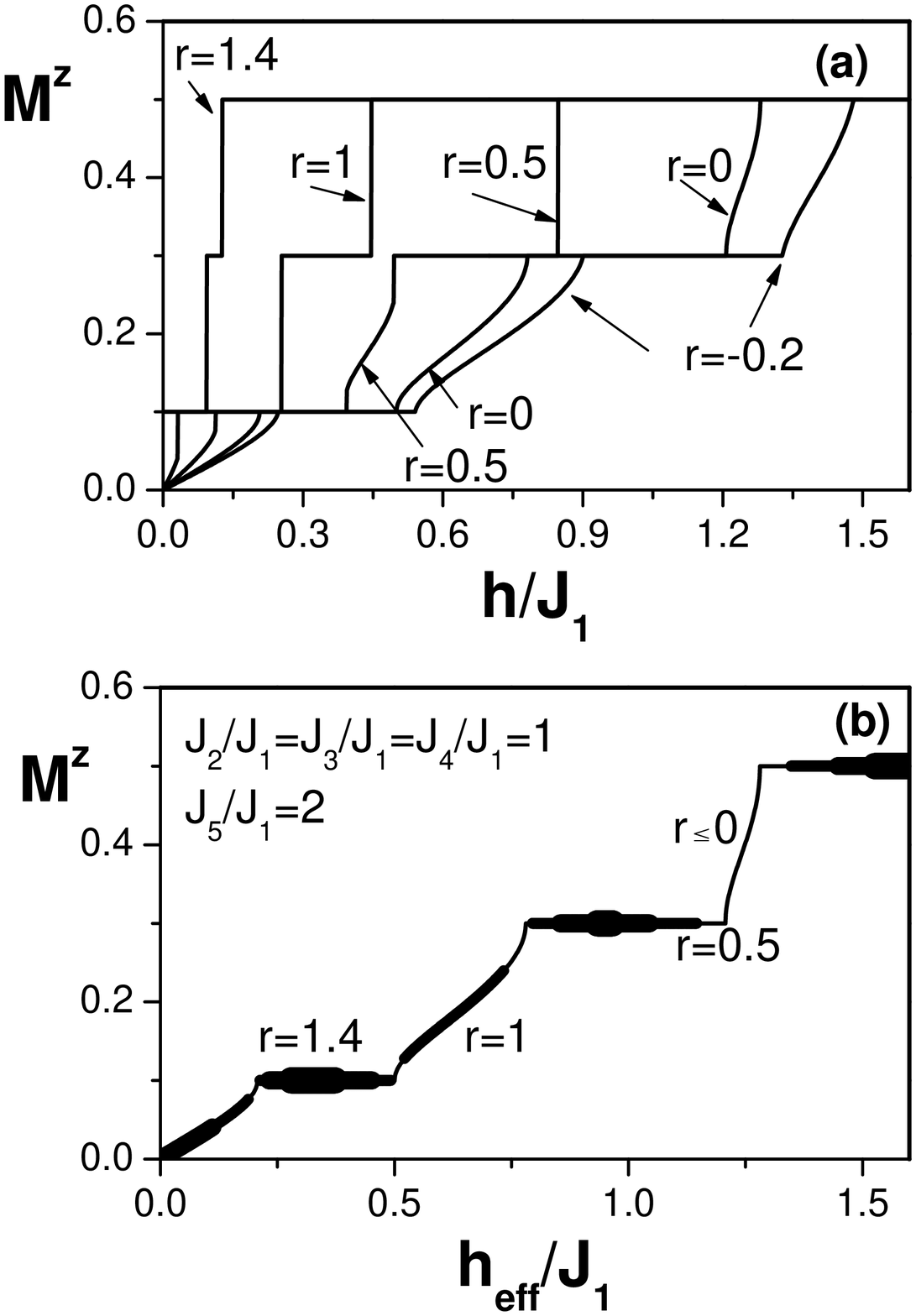}%
\caption{(a) Magnetization as a function of $h/J_{1}$, and (b) universal curve
for the magnetization as a function of the effective field $h_{eff}/J_{1}$
$(h_{eff}=h+2J^{\prime}M^{z}),$ at $T=0,$ for different values of
$r$($r=J^{\prime}/J_{1})$, in the regions where the system undergoes first
($r>0)$ and second-order ($r\leqslant0)$ quantum transitions, for $n=5$ and
$J_{1}=1,$ $J_{2}=1,J_{3}=1,J_{4}=1,J_{5}=2.$}%
\end{center}
\end{figure}

%

\begin{figure}
[ptb]
\begin{center}
\includegraphics[
trim=0.000000in 0.000000in 0.000000in 1.830541in,
height=5.9127in,
width=3.3434in
]%
{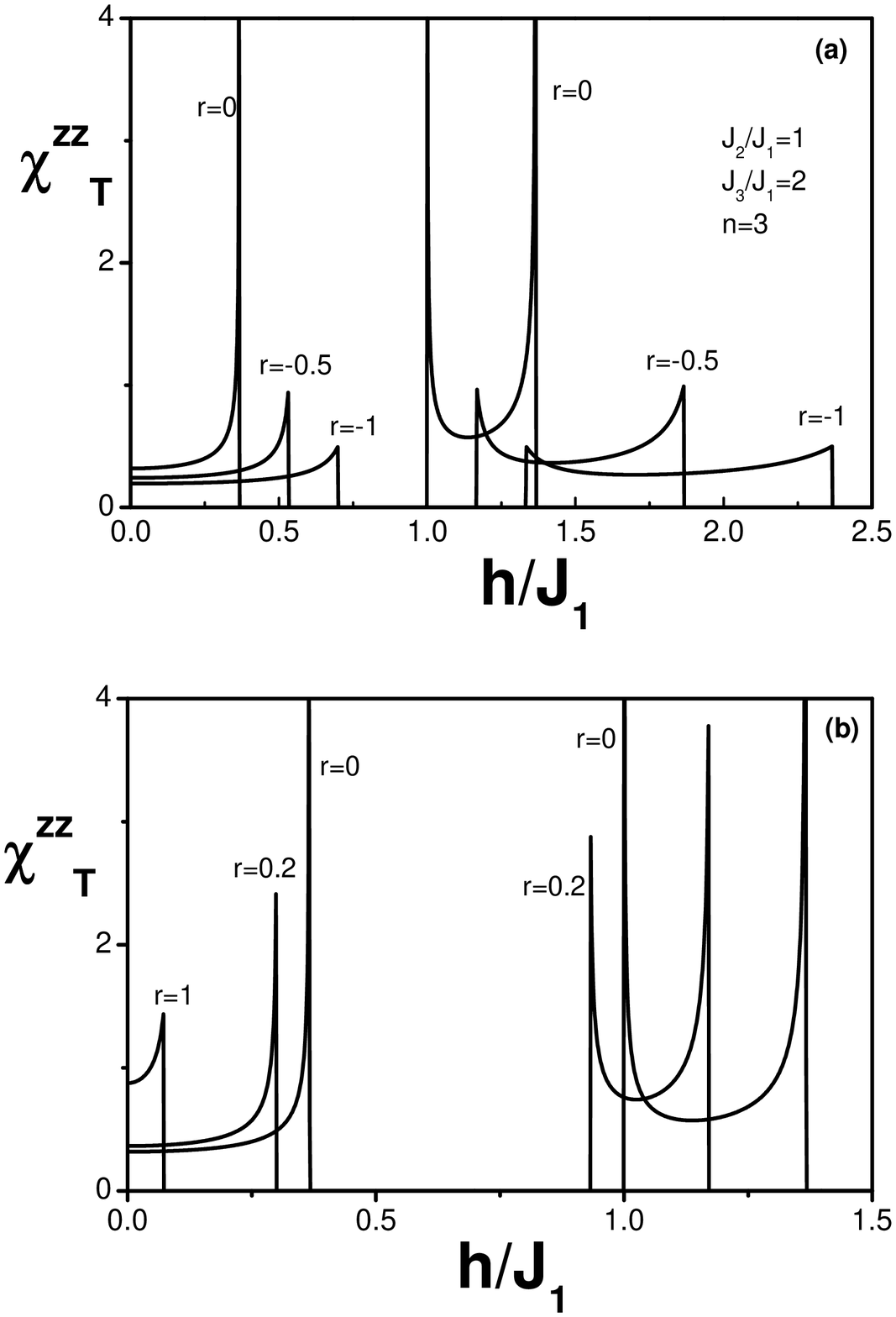}%
\caption{Isothermal susceptibility $\chi_{T}^{zz},$at $T=0,$ as a function of
$h/J_{1}$, for $n=3,$ $J_{1}=1,J_{2}=1,$ $J_{3}=2,$ and different values of
$r$($r=J^{\prime}/J_{1})$, (a) for $r\leqslant0$ and (b) for $r\geqslant0.$}%
\end{center}
\end{figure}

\section{Acknowledgements}

The authors would like thank the Brazilian agencies CNPq and Capes for partial
financial support.

\end{document}